\documentstyle[psfig,conf_iap,10pt]{article}

\begin{document}

\heading{Galaxy Counterparts of High-Redshift DLA Systems}
 
\par\medskip\noindent

\author{S. G. Djorgovski}

\address{Palomar Observatory, Caltech, Pasadena, CA 91125, USA}

\begin{abstract}
Our understanding of the nature of DLA systems at large redshifts, ostensibly
progenitors of normal disk galaxies, depends critically on their direct
identifications with galaxies, and the resulting measurements of their
properties.  A few such objects have now been found, reaching out to $z \approx
4.1$.  Their observed luminosities are $L \sim L_*$, star formation rates $SFR
\sim$ a few $M_\odot$/yr, physical sizes $\sim 20$ kpc, and velocity fields of
a few hundred km/s, implying masses $> 10^{11} M_\odot$.  While their
morphology remains uncertain, the observed properties are consistent with those
expected of young disk galaxies in the early stages of formation.  We also find
a statistically significant excess of foreground galaxies near lines of sight
to luminous quasars at $z > 4$.  This suggests a systematical gravitational
lensing magnification of such quasar samples, possibly with important
consequences for the estimates of the quasar luminosity function at high
redshifts, and the deduced $\Omega_b$ in DLA systems found in their spectra. 

\end{abstract}

\section{ Introduction }

Galaxies believed to be responsible for damped Ly$\alpha$ absorption (DLA)
systems in the spectra of high-redshift quasars represent a viable population
of progenitors of normal disk galaxies \cite{Wolfe1993}.  
It is also possible that they represent still merging, gas-rich protogalactic
clumps, rather than already formed disks.  They appear to contain a substantial
fraction of the baryons known to exist in normal galaxies 
today \cite{StorLom1996}.
DLA systems represent an already well-established, {\it large population} of
high-$z$ objects for which the confusion with AGN does not arise.

The crucial question is whether DLA systems represent a population of already
assembled massive proto-disks, as advocated, e.g., by Wolfe and collaborators,
or whether they are still subgalactic fragments in the process of hierarchical
assembly, as favored by many n-body simulators \cite{Katz1996}. 
In order to answer this question, it is essential to obtain firm optical
identifications of objects responsible for the DLA absorption, and to determine
their physical properties. 

Direct imaging of DLA galaxies can provide measurements of their luminosities. 
From their separation from the QSO line of sight, one can infer their physical
sizes.  Coupled with the spectroscopic measurements of velocity fields in these
systems, one can deduce their dynamical masses.  Sizes and H I column densities
give the gas masss.  Star formation rates can be obtained from the UV continuum
luminosity, and from the Ly$\alpha$ line emission, if present. 

Clustered companions of DLA systems, all containing AGN, or quasar companions
which may be responsible for some associated absorption have been 
seen in several cases \cite{Lowe1991,Macc1993,Malkan1995}, 
but until recently no normal, isolated DLA systems themselves.  Warren \&
Moller \cite{WarMol1996}
found emission-line regions associated with a radio-loud quasar PKS 0528--250
at $z = 2.81$, which also has an associated DLA system.  The relation of these
emission-line objects with the absorber is not clear; yet Ly$\alpha$ companions
of radio-loud quasars have been seen in many cases.  The whole point of
studying DLA galaxies is to get away from systems dominated by AGN, by
selecting isolated, field objects.  Even if there are DLA systems associated
with quasars, their utility for the purpose of understanding of formation of
normal disk galaxies is rather uncertain.

\section{ Detections of DLA Galaxies at High Redshifts }

We discovered for the first time a galaxy responsible for a known, isolated DLA
system at $z_{abs} = 3.150$ in the spectrum of a quasar 
2233+131 \cite{Djorg1996}.  
Its observed physical properties and the derived SFR correspond closely to 
what may be expected of a young disk galaxy still in the process of formation: 
no spectroscopic evidence for an AGN, luminosity $\sim L_*$, inferred star
formation rate (both from the Ly$\alpha$ line and the UV continuum) SFR
$\approx 7 M_\odot$/yr, a physical size (from the separation of the absorber
portion in the front of the QSO and the counterpart galaxy emission) of $\sim
15$ kpc in the restframe, and a velocity field with an amplitude of $\sim 200$
km/s.  These properties are exactly what may be expected from a young, massive
disk galaxy. 

\begin{figure}
\centerline{\vbox{
\psfig{figure=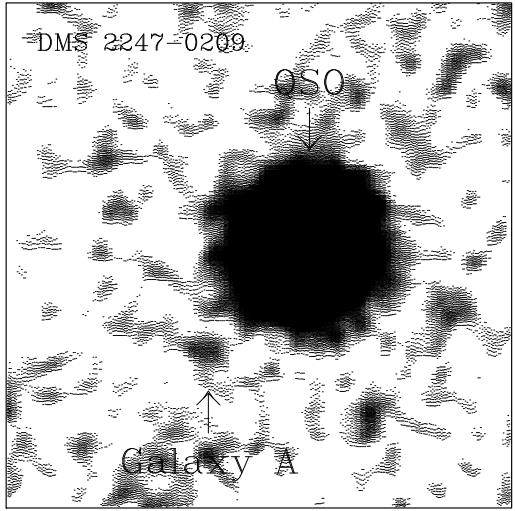,height=7.0cm,angle=0}
}}
\caption[]{
Deep $R$ band Keck image of the field of DMS 2247--0209, a quasar at $z =
4.36$.  Galaxy A is an $R \approx 26^m$ object at the redshift of a DLA system
at $z = 4.10$ seen in the quasar spectrum.  Image shown is 11 arcsec square.
}
\end{figure}

Since then, we have stared a systematic search for more such high-$z$ DLA
counterparts at the W.M. Keck telescope.  So far, we have found a several
candidates, most of which still require more spectroscopy.  Typically, these
galaxies have $R > 24^m$ and no strong line emission, in agreement with the
results found for the field Lyman-break galaxies \cite{Steidel1996}. 

The most interesting case so far (Figure 1) is a possible counterpart of a DLA
system at $z = 4.1$, seen in the spectrum of the quasar DMS 2247--0209
($z_{QSO} = 4.36$) \cite{Hall1996}. 
The object is an $R \approx 26^m$ galaxy seen 3.3 arcsec in projection from the
QSO l.o.s. (corresponding to $\sim 110$ comoving kpc, or $\sim 22$ proper kpc
at the absorber redshift, for $h \sim 0.7$ and $\Omega_0 \sim 0.2$).  Weak
Ly$\alpha$ emission line (Figure 2) is detected at $z_{gal} = 4.097$.  Its
inferred continuum luminosity is $\sim 0.5 L_*$, and the SFR $\approx 0.7
M_\odot$/yr.  In other words, we can now study nearly dwarf galaxies at $z >
4$!  Of course, its low luminosity and SFR may be indicative of its extreme
youth, rather than its mass. 

\begin{figure}
\centerline{\vbox{
\psfig{figure=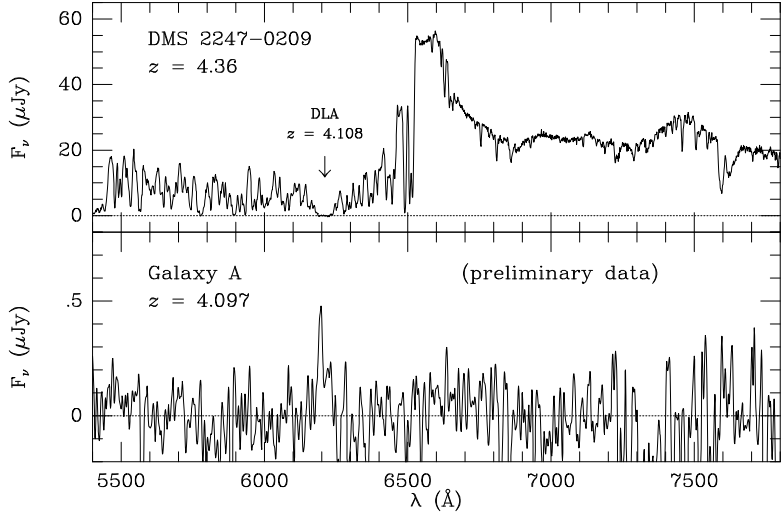,height=7.0cm,angle=0}
}}
\caption[]{
Preliminary Keck spectra of the quasar DMS 2247--0209, and the galaxy A,
associated with the DLA system at $z = 4.10$.  The glitch to the immediate
right of the Ly$\alpha$ line is an artifact of the sky subtraction. 
}
\end{figure}

To summarize, direct galaxy counterparts of DLA systems at $z \sim 3 - 4$ have
been found.  Their typical magnitudes are $R \sim 25^m$, with separations from
the QSO l.o.s. $\sim 2 - 3$ arcsec (the inner limit is due to the seeing). For
a reasonable range of cosmological parameters, the corresponding luminosities
are $L \sim L_*$, star formation rates (determined from both the UV coninuum
luminosity and from the Ly$\alpha$ emission line) are $SFR \sim$ a few
$M_\odot$/yr, and the projected physical sizes are $\sim 20$ kpc.  Along with
the typical observed velocity fields of $\sim 200 - 300$ km/s, this implies
dynamical masses $> 10^{11} M_\odot$. 

\section{ Concluding Comments }

These preliminary results support the idea that DLA systems represent likely
progenitors of normal disk galaxies today.  The populations of DLA systems, of
which DLA 2233+131 and DLA 2247--021 may be representative, and of Lyman-break
objects studied by Steidel and others \cite{Steidel1996}
may overlap considerably, and they may be tentatively identified as progenitors
of typical normal (disk?) galaxies today. 

Intriguingly, we have found a significant excess (a 7-$\sigma$ effect) of
apparently foreground galaxies near the lines of sight to luminous quasars at
$z > 4$, which may be also correlated with the apparent luminosity of the QSO 
\cite{Djorg1997}.
Our spectroscopy of these galaxies so far shows that nearly all are foreground
objects, typically at $z \sim 1$, although at least a few are quasar companions
at $z > 4$ \cite{Djorg1997}.  This suggests that gravitational (micro)lensing
may be systematically affecting our inferred luminosities of these objects. 

This result may have significant implications for our understanding of the
evolution of the quasar luminosity function \cite{Kenn1995}, as well as the
evolution of the high-redshift absorbers seen in their spectra
\cite{StorLom1996}: both the abundance of high-luminosity quasars and the 
comoving density of H I in DLA's at high redshifts would be overestimated. 
However, if a population of optically thick lenses/absorbers exists, that would
have the opposite effect \cite{Fall1993}.

\acknowledgements{ 
I wish to thank my collaborators, especially M.~Pahre, R.~Gal, J.~Kennefick,
and R.~de Carvalho, to the staff of W.M. Keck Observatory for the expert help
during numerous observing runs, and to the organizers of this conference for
their efforts.  This work was supported in part by the NSF PYI award
AST-9157412 and by the Bressler Foundation. 
}

\end{document}